\documentstyle{article}
\newtheorem{rem}{Remark}

\begin{document}
\begin{center}
\Large{\bf ON THE RIEMANN EXTENSION OF \\[2mm]THE G\"ODEL
SPACE-TIME METRIC } \vspace{4mm}\normalsize
\end{center}
 \begin{center}
\Large{\bf Valery Dryuma}\vspace{4mm}\normalsize
\end{center}
\begin{center}
{\bf Institute of Mathematics and Informatics AS Moldova, Kishinev}\vspace{4mm}\normalsize
\end{center}
\begin{center}
{\bf  E-mail: valery@dryuma.com;\quad cainar@mail.md \quad dryuma@math.md}\vspace{4mm}\normalsize
\end{center}
\begin{center}
{\bf  Abstract}\vspace{4mm}\normalsize
\end{center}

   Some properties of the  G\"odel space-metric and its Riemann extension are
   studied.\\[0.5cm]

 \hfill         SATOR AREPO TENET OPERA ROTAS

\section{Introduction}

   The notion of the Riemann extension of nonriemannian spaces was introduced first in
    ~(\cite{dryuma1:paterson&walker}).
Main idea of this theory is application of the methods of Riemann
geometry for studying of the properties of nonriemaniann spaces.

For example the system differential equations in form
\begin{equation} \label{dryuma:eq1}
\frac{d^2 x^{k}}{ds^2}+\Pi^k_{ij}\frac{dx^{i}}{ds}\frac{dx^{j}}{ds}=0
\end{equation}
with arbitrary coefficients $\Pi^k_{ij}(x^l)$ can be considered as the system of geodesic equations of
affinely connected space with local coordinates $x^k$.

 For the n-dimensional Riemannian spaces with the metrics
\[
{^n}ds^2=g_{ij}dx^i dx^j
\] the system of geodesic equations looks
similar but the coefficients $\Pi^k_{ij}(x^l)$ now have very
special form and depends from the choice of the metric $g_{ij}$.
\[
\Pi^i_{kl}=\Gamma^i_{kl}=\frac{1}{2}g^{im}(g_{mk,l}+g_{ml,k}-g_{kl,m})
\]

In order that methods of Riemann geometry can be applied for studying of the properties of the spaces with
equations (\ref{dryuma:eq1}) the construction of 2n-dimensional extension of the space with local coordinates $x^i$  was introduced .

The metric of extended space constructs with help of coefficients of equation (\ref{dryuma:eq1}) and looks as follows
\begin{equation} \label{dryuma:eq2}
{^{2n}}ds^2=-2\Pi^k_{ij}(x^l)\Psi_k dx^i dx^j+2d \Psi_k dx^k
\end{equation}
where $\Psi_{k}$ are the coordinates of additional space.

   The important property of such type metric is that the geodesic
 equations of metric (\ref{dryuma:eq2})  decomposes into two parts
\begin{equation} \label{dryuma:eq3}
\ddot x^k +\Gamma^k_{ij}\dot x^i \dot x^j=0,
\end{equation}
and
\begin{equation} \label{dryuma:eq4}
\frac{\delta^2 \Psi_k}{ds^2}+R^l_{kji}\dot x^j \dot x^i \Psi_l=0,
\end{equation}
where
\[
\frac{\delta \Psi_k}{ds}=\frac{d \Psi_k}{ds}-\Gamma^l_{jk}\Psi_l\frac{d x^j}{ds}.
\]

The first part (\ref{dryuma:eq3}) of the full system is the system
of equations for geodesics of basic space with local coordinates
$x^i$ and it does not contains the coordinates $\Psi_k$.

 The second part (\ref{dryuma:eq4}) of system of geodesic equations  has the form
of linear $4\times4$ matrix system of second order ODE's for coordinates $\Psi_k$
\begin{equation} \label{dryuma:eq5}
\frac{d^2 \vec \Psi}{ds^2}+A(s)\frac{d \vec \Psi}{ds}+B(s)\vec \Psi=0
\end{equation}
with the matrix
\[
A=A(x^i(s),\dot x^i(s)), \quad B=B(x^i(s),\dot x^i(s)).
\]

From this point of view we have the case of geodesic extension of
the basic space  $(x^i)$.

  It is important to note that the geometry of extended space is connected with
geometry of basic space.

   For example the property of  the space to be a
   Ricci-flat $$R_{ij}=0,$$ or $$R_{i j;k}+
R_{k i;j}+R_{j k;i}=0,$$ or symmetrical $$R_{i j k l;m}=0$$ keeps also for the
extended space.

     This fact give us the possibility to use the linear system of equation (\ref{dryuma:eq5}) for studying
of the properties of basic space.

  In particular the invariants of $4\times4$
matrix-function
\[
E=B-\frac{1}{2}\frac{d A}{ds}-\frac{1}{4}A^2
\]
under change of the coordinates $\Psi_k$ can be of used for that.

    For example the condition
\[
E=B-\frac{1}{2}\frac{d A}{ds}-\frac{1}{4}A^2=0
\]
for a given system means that it is equivalent to the simplest system
\[
\frac{d^2 \vec \Phi}{ds^2}=0
\]
and corresponding extended space is a flat space.

    Other cases of integrability of the system ~(\ref{dryuma:eq5})
    are connected with a not flat spaces having special form of the curvature tensor.

   Remark that for  extended spaces all scalar invariants constructed with
   the help of curvature tensor and its covariant derivatives are vanishing.

   The first applications of the notion of extended spaces to the studying of nonlinear second order differential
 equations and the Einstein spaces were done in the works of author
~(\cite{dryuma2:dryuma})-(\cite{dryuma11:dryuma}).

 Here we consider the properties of the G\"odel space-time and its Riemann
 extension.

\section{The G\"odel space-time metric}
The line element of the metric of the G\"odel space-time in
coordinates $x ,y,z,t$ has the form

\begin{equation} \label{dryuma:eq6}
ds^2=-{d{{t}}}^{2}+{d{{x}}}^{2}-2\,{e^{{\frac {x}{a}}}}d{{t}}d{{y}}-1/2
\,{e^{2\,{\frac {x}{a}}}}{d{{y}}}^{2}+{d{{z}}}^{2} .
\end{equation}

   Here the parameter $a$ is the velocity of rotation
   ~(\cite{dryuma12:hawking}).

The geodesic equations of the metric ~(\ref{dryuma:eq6}) are given by
\begin{equation} \label{dryuma:eq7}
2\, \left( {\frac {d^{2}}{d{s}^{2}}}x \left( s \right)  \right) a+
 \left( {e^{{\frac {x \left( s \right) }{a}}}} \right) ^{2} \left( {
\frac {d}{ds}}y \left( s \right)  \right) ^{2}+2\,{e^{{\frac {x
 \left( s \right) }{a}}}} \left( {\frac {d}{ds}}t \left( s \right)
 \right) {\frac {d}{ds}}y \left( s \right)
=0,
\end{equation}
\begin{equation} \label{dryuma:eq8}
 \left( {\frac {d^{2}}{d{s}^{2}}}y \left( s \right)  \right) {e^{{
\frac {x \left( s \right) }{a}}}}a-2\, \left( {\frac {d}{ds}}t \left( s \right)
\right) {\frac {d}{ds}}x \left( s \right) =0,
\end{equation}
\begin{equation}\label{dryuma:eq9}
{\frac {d^{2}}{d{s}^{2}}}z(s)=0,
\end{equation}
\begin{equation} \label{dryuma:eq10}
 \left( {\frac {d^{2}}{d{s}^{2}}}t \left( s \right)  \right) a+{e^{{
\frac {x \left( s \right) }{a}}}} \left( {\frac {d}{ds}}y \left( s
 \right)  \right) {\frac {d}{ds}}x \left( s \right) +2\, \left( {
\frac {d}{ds}}t \left( s \right)  \right) {\frac {d}{ds}}x \left( s
 \right)
=0.
\end{equation}

   The first integral of geodesics is satisfied  the condition
$$- \left( {\frac {d}{ds}}t \left( s \right)  \right) ^{2}+ \left( { \frac {d}{ds}}x
\left( s \right)  \right) ^{2}-2\,{e^{{\frac {x
 \left( s \right) }{a}}}} \left( {\frac {d}{ds}}t \left( s \right)
 \right) {\frac {d}{ds}}y \left( s \right) -1/2\,{e^{2\,{\frac {x
 \left( s \right) }{a}}}} \left( {\frac {d}{ds}}y \left( s \right)
 \right) ^{2}+\]\[+ \left( {\frac {d}{ds}}z \left( s \right)  \right) ^{2}-
\mu=0. $$

 The symbols of Christoffel of the metric
~(\ref{dryuma:eq6}) are
 $$
  \Gamma^4_{12}=\frac{\exp(x/a)}{2a}
 ,\quad
\Gamma^2_{14}=-\frac{exp(-x/a)}{a},\quad \Gamma^4_{14}=\frac{1}{a},\quad
\Gamma^1_{22}=\frac{exp(2x/a)}{2a},\quad \Gamma^1_{24}=\frac{exp(x/a)}{2a}.
$$

   To find the solutions of the equations of geodesics
(\ref{dryuma:eq7}--\ref{dryuma:eq10}) we present the metric
~(\ref{dryuma:eq6}) in equivalent form
~(\cite{dryuma13:dautcourt})
\begin{equation}\label{dryuma:eq11}
ds^2=-(dt+\frac{a
\sqrt{2}}{y}dx)^2+\frac{a^2}{y^2}(dx^2+dy^2)+dz^2.
\end{equation}

 The  correspondence between the both forms of the metrics is given by the relations
 $$
 y=a \sqrt{2}\exp(-x/a),\quad x=y.
 $$

 The equations of geodesics of the metric (\ref{dryuma:eq11}) are
 defined by
\begin{equation}\label{dryuma:eq12}
\left ({\frac {d^{2}}{d{s}^{2}}}x(s)\right )a+\sqrt {2}\left
({\frac { d}{ds}}t(s)\right ){\frac {d}{ds}}y(s) =0,
\end{equation}
\begin{equation}\label{dryuma:eq13}
\left ({\frac {d^{2}}{d{s}^{2}}}y(s)\right )y(s)a-\left ({\frac
{d}{ds }}x(s)\right )^{2}a-\sqrt {2}\left ({\frac
{d}{ds}}t(s)\right )\left ( {\frac {d}{ds}}x(s)\right )y(s)-\left
({\frac {d}{ds}}y(s)\right )^{2} a  =0,
\end{equation}
\begin{equation}\label{dryuma:eq14}
\left ({\frac {d^{2}}{d{s}^{2}}}t(s)\right )\left (y(s)\right
)^{2}- \sqrt {2}a\left ({\frac {d}{ds}}y(s)\right ){\frac
{d}{ds}}x(s)-2\, \left ({\frac {d}{ds}}t(s)\right )\left ({\frac
{d}{ds}}y(s)\right )y( s) =0,
\end{equation}
\begin{equation}\label{dryuma:eq15}
{\frac {d^{2}}{d{s}^{2}}}z(s) =0.
\end{equation}

 The geodesic equations admit the first integral
\begin{equation}\label{dryuma:eq16}
 \frac{dt}{ds}=\frac{(-c_2/\sqrt{2}+\sqrt{2}y)}{c_0},
$$ $$ \frac{dx}{ds}=\frac{y(c_2-y)}{a c_0}, $$ $$
\frac{dy}{ds}=\frac{y(x-c_1)}{a c_0}, $$ $$
\frac{dz}{ds}=\frac{c_3}{c_0}, \end{equation}
 where $c_i,a_0$ are the parameters.

\begin{rem}

 In theory of varieties the Chern-Simons characteristic class is constructed from a
 matrix gauge connection $A^i_{jk}$ as
\[
W(A)=\frac{1}{4\pi^2}\int d^3x \epsilon^{ijk}tr\left(\frac{1}{2}A_i\partial_j
A_k+\frac{1}{3}A_iA_jA_k \right).
\]

   This term can be translated into a three-dimensional geometric quantity by
   replacing the matrix connection $A^i_{jk}$ with the Christoffel connection $\Gamma^i_{jk}$.

   For the density of  Chern-Simons invariant can be obtained the
   expression (\cite{dryuma14:jac})
$$ CS(\Gamma)=\epsilon^{i j k}(\Gamma^p_{i
q}\Gamma^q_{k p;j}+\frac{2}{3}\Gamma^p_{i q}\Gamma^q_{j r}\Gamma^r_{k p}).
$$

   For the metric (\ref{dryuma:eq11}) at the condition $z=const$
$$
ds^2=-a^2/y^2 dx^2 -2 \sqrt{2} a/y dxdt+a^2/y^2 dy^2- dt^2
$$
    we find the quantity
\[
CS(\Gamma)=-{\frac {\sqrt {2}}{a{y}^{2}}} .
\]

For the spatial metric
$$
^{3}ds^2=-g_{\alpha \beta}+\frac{g_{0 \alpha}g_{0 \beta}}{g_{0 0}}
$$
 of the metric (\ref{dryuma:eq11})
$$
-ds^2=\frac{a^2}{y^2}(dx^2 +dy^2)+ dz^2
$$
 the quantity
\[
CS(\Gamma)=0.
\]

\end{rem}

\section{ The Riemann extension of the G\"odel metric}

    The Christofell symbols of the metric (\ref{dryuma:eq11}) are
\[
\Gamma^2_{11}=-\frac{1}{y}, \quad \Gamma^2_{22}=-\frac{1}{y},
\quad \Gamma^2_{14}=-\frac{\sqrt{2}}{2a}, \quad
\Gamma^1_{24}=\frac{\sqrt{2}}{2a},\quad
\Gamma^4_{12}=-\frac{a\sqrt{2}}{2y^2},\quad
\Gamma^4_{14}=-\frac{1}{y}.
\]

        Now with help of the formulae (\ref{dryuma:eq2}) we construct the eight-dimensional
extension of the metric (\ref{dryuma:eq11}).

 It has the form
\begin{equation}\label{dryuma:eq17}
^{8}ds^2=\frac{2}{y}Q dx^2+\frac{2 a \sqrt{2}}{y^2}V dx dy+\frac{2
\sqrt{2}}{a}Q dx dt+\frac{2}{y} Q dy^2+(\frac{4}{y} V -2
\frac{\sqrt{2}}{a}P)dy dt+\]\[+2d x dP+ 2d y d Q +2 d z dU +2 dt
dV.
\end{equation}
where $(P,Q,U,V)$ are an additional coordinates.

   The Ricci tensor of the four-dimensional G\"odel space with the metric (\ref{dryuma:eq11}) or (\ref{dryuma:eq6})
satisfied the condition
\[
{^4}R_{i k;l}+{^4}R_{l i;k}+{^4}R_{k l;i}=0.
\]

    This property is valid for the eight-dimensional space in local coordinates $(x,y,z, t,P,Q,U,V)$
 with the metric (\ref{dryuma:eq11})
\[
{^8}R_{i k;l}+{^8}R_{l i;k}+{^8}R_{k l;i}=0.
\]

  The full system of  geodesic equations for the metric (\ref{dryuma:eq7})  decomposes into two parts.

The first part coincides with the equations
(\ref{dryuma:eq12}-\ref{dryuma:eq15}) on the coordinates $(x,y,z,
t)$ and second  part forms the linear system of equations for
coordinates $P,Q,U,V$.

 They are defined as
\begin{equation}\label{dryuma:eq18}
{\frac {d^{2}}{d{s}^{2}}}P(s)=-{\frac {\left (\sqrt
{2}{a}^{2}\left ({ \frac {d}{ds}}x(s)\right )^{2}+2\,\left ({\frac
{d}{ds}}t(s)\right ) \left ({\frac {d}{ds}}x(s)\right )y(s)a-\sqrt
{2}{a}^{2}\left ({\frac {d}{ds}}y(s)\right )^{2}\right
)V(s)}{\left (y(s)\right )^{3}a}}-\]\[-{ \frac {\left (2\,\left
({\frac {d}{ds}}x(s)\right )\left (y(s)\right ) ^{2}a+\sqrt
{2}\left ({\frac {d}{ds}}t(s)\right )\left (y(s)\right )^{
3}\right ){\frac {d}{ds}}Q(s)}{\left (y(s)\right )^{3}a}}-{\frac {
\sqrt {2}a\left ({\frac {d}{ds}}V(s)\right ){\frac {d}{ds}}y(s)}{
\left (y(s)\right )^{2}}},
\end{equation}
\\[2mm]
\begin{equation}\label{dryuma:eq19}
{\frac {d^{2}}{d{s}^{2}}}Q(s)={\frac {\left (-3\,\left ({\frac
{d}{ds} }x(s)\right )^{2}y(s)a-2\,\sqrt {2}\left ({\frac
{d}{ds}}t(s)\right ) \left ({\frac {d}{ds}}x(s)\right )\left
(y(s)\right )^{2}-\left ({ \frac {d}{ds}}y(s)\right
)^{2}y(s)a\right )Q(s)}{\left (y(s)\right )^{ 3}a}}+\]\[+{\frac
{\left (2\,a\left ({\frac {d}{ds}}y(s)\right )\left ({ \frac
{d}{ds}}x(s)\right )y(s)+2\,\left ({\frac {d}{ds}}t(s)\right )
\left ({\frac {d}{ds}}y(s)\right )\left (y(s)\right )^{2}\sqrt {2}
\right )P(s)}{\left (y(s)\right )^{3}a}}+\]\[+{\frac {\left
(-2\,\left ({ \frac {d}{ds}}t(s)\right )\left ({\frac
{d}{ds}}y(s)\right )y(s)a-2\, \sqrt {2}{a}^{2}\left ({\frac
{d}{ds}}y(s)\right ){\frac {d}{ds}}x(s) \right )V(s)}{\left
(y(s)\right )^{3}a}}+{\frac {\sqrt {2}\left ({ \frac
{d}{ds}}t(s)\right ){\frac {d}{ds}}P(s)}{a}}-\]\[-2\,{\frac {\left
({ \frac {d}{ds}}y(s)\right ){\frac {d}{ds}}Q(s)}{y(s)}}+{\frac
{\left (- \sqrt {2}{a}^{2}\left ({\frac {d}{ds}}x(s)\right
)y(s)-2\,\left ({ \frac {d}{ds}}t(s)\right )\left (y(s)\right
)^{2}a\right ){\frac {d}{d s}}V(s)}{\left (y(s)\right )^{3}a}},
\end{equation}
\\[2mm]
\begin{equation}\label{dryuma:eq20}
{\frac {d^{2}}{d{s}^{2}}}U(s)=0,
\end{equation}
\\[2mm]
\begin{equation}\label{dryuma:eq21}
{\frac {d^{2}}{d{s}^{2}}}V(s)={\frac {\left (\left ({\frac
{d}{ds}}x(s )\right )^{2}a\sqrt {2}y(s)+2\,\left ({\frac
{d}{ds}}t(s)\right ) \left ({\frac {d}{ds}}x(s)\right )\left
(y(s)\right )^{2}+\sqrt {2} \left ({\frac {d}{ds}}y(s)\right
)^{2}ay(s)\right )P(s)}{{a}^{2}\left (y(s)\right
)^{2}}}+\]\[+{\frac {\left (-2\,\sqrt {2}\left ({\frac {d}{ds}}
t(s)\right )\left ({\frac {d}{ds}}x(s)\right
)y(s)a-2\,{a}^{2}\left ({ \frac {d}{ds}}x(s)\right )^{2}\right
)V(s)}{{a}^{2}\left (y(s)\right ) ^{2}}}+{\frac {\sqrt {2}\left
({\frac {d}{ds}}y(s)\right ){\frac {d}{d s}}P(s)}{a}}-\]\[-{\frac
{\sqrt {2}\left ({\frac {d}{ds}}x(s)\right ){ \frac
{d}{ds}}Q(s)}{a}}-2\,{\frac {\left ({\frac {d}{ds}}y(s)\right ){
\frac {d}{ds}}V(s)}{y(s)}}+2\,{\frac {Q(s)\left ({\frac
{d}{ds}}t(s) \right ){\frac {d}{ds}}y(s)}{{a}^{2}}}.
\end{equation}

     In result we have got a linear  matrix-second order ODE for the coordinates $U,V,P,Q$
\begin{equation}\label{dryuma:eq22}
\frac{d^2\Psi}{ds^2}=A(x,\phi,z,t)\frac{d\Psi}{ds}+B(x,\phi,z,t)\Psi,
\end{equation}
where
\[
\Psi(s)=\left(\begin{array}{c}
P(s)\\
Q(s)\\
U(s)\\
V(s)
\end{array}\right)
\]
and $A,B$ are some $4 \times 4$ matrix-functions depending from
the coordinates $x(s),y(s),z(s),t(s)$ and their derivatives.

    Now we shall investigate the properties of the matrix system of equations
     (\ref{dryuma:eq18}-\ref{dryuma:eq21}).

    To integrate this system we use the relation
\begin{equation}\label{dryuma:eq23}
\dot x(s) P(s)+\dot y(s) Q(s)+\dot z(s) U(s)+ \dot t(s)
V(s)-\frac{s}{2}-\mu=0,
\end{equation}
which is valid for the every Riemann extensions of affinely
connected space and which is consequence of the well known first
integral of geodesic equations
\[
g_{ik}\dot x^{i}\dot x^{k}=\nu
\]
 of arbitrary Riemann space.

    Using the expressions for the first integrals of geodesic (\ref{dryuma:eq16}) and
$
 U(s)=\alpha s+\beta
$
 from the equation (\ref{dryuma:eq20})   the system of equations
(\ref{dryuma:eq18}-\ref{dryuma:eq21}) may be simplified.

    In result we get the system of equations for additional coordinates
\begin{equation}\label{dryuma:eq24}
{\frac {d^{2}}{d{s}^{2}}}P \left( s \right) ={\frac { \left( \sqrt {2} {\it
c0}\,a{\it c1}-\sqrt {2}{\it c0}\,ax \left( s \right)  \right) { \frac {d}{ds}}V
\left( s \right) }{y \left( s \right) {{\it c0}}^{2}a} }+\]\[+{\frac { \left( -\sqrt
{2}{\it c2}\,y \left( s \right) -2\,\sqrt {2} x \left( s \right) {\it c1}+\sqrt
{2}{{\it c1}}^{2}+\sqrt {2} \left( y
 \left( s \right)  \right) ^{2}+\sqrt {2} \left( x \left( s \right)
 \right) ^{2} \right) V \left( s \right) }{y \left( s \right) {{\it c0
}}^{2}a}}-{\frac { \left( {\frac {d}{ds}}Q \left( s \right)  \right) { \it c2}}{{\it
c0}\,a}}  ,
\end{equation}
\\[2mm]
\begin{equation}\label{dryuma:eq25}
{\frac {d^{2}}{d{s}^{2}}}Q \left( s \right) =-{\frac { \left( 2\,
 \left( y \left( s \right)  \right) ^{2}{\it c1}-2\, \left( y \left( s
 \right)  \right) ^{2}x \left( s \right)  \right) P \left( s \right) }
{y \left( s \right) {a}^{2}{{\it c0}}^{2}}}-\]\[-{\frac { \left( y \left( s
 \right)  \left( x \left( s \right)  \right) ^{2}- \left( y \left( s
 \right)  \right) ^{3}+y \left( s \right) {{\it c1}}^{2}+y \left( s
 \right) {{\it c2}}^{2}-2\,y \left( s \right) x \left( s \right) {\it
c1} \right) Q \left( s \right) }{y \left( s \right) {a}^{2}{{\it c0}}^
{2}}}-\]\[-{\frac { \left( 2\,a{\it c0}\,y \left( s \right) x \left( s
 \right) -2\,a{\it c0}\,y \left( s \right) {\it c1} \right) {\frac {d}
{ds}}Q \left( s \right) }{y \left( s \right) {a}^{2}{{\it c0}}^{2}}}-{ \frac {
\left( a{\it c0}\,y \left( s \right) {\it c2}-2\,a{\it c0}\,
 \left( y \left( s \right)  \right) ^{2} \right) {\frac {d}{ds}}P
 \left( s \right) }{y \left( s \right) {a}^{2}{{\it c0}}^{2}}}-\]\[-{\frac
{\sqrt {2}{\frac {d}{ds}}V \left( s \right) }{{\it c0}}}-{\frac {
 \left( \sqrt {2}{\it c2}\,ax \left( s \right) -\sqrt {2}{\it c2}\,a{
\it c1} \right) V \left( s \right) }{y \left( s \right) {a}^{2}{{\it c0}}^{2}}},
\end{equation}
\\[2mm]
\begin{equation}\label{dryuma:eq26}
{\frac {d^{2}}{d{s}^{2}}}V \left( s \right) ={\frac { \left(  \left( y
 \left( s \right)  \right) ^{2}\sqrt {2}{\it c2}- \left( y \left( s
 \right)  \right) ^{3}\sqrt {2}+y \left( s \right) \sqrt {2} \left( x
 \left( s \right)  \right) ^{2}-2\,y \left( s \right) \sqrt {2}x
 \left( s \right) {\it c1}+y \left( s \right) \sqrt {2}{{\it c1}}^{2}
 \right) P \left( s \right) }{{{\it c0}}^{2}{a}^{3}}}+\]\[+{\frac { \left(
-\sqrt {2}y \left( s \right) x \left( s \right) {\it c2}+\sqrt {2}y
 \left( s \right) {\it c1}\,{\it c2}+2\,\sqrt {2} \left( y \left( s
 \right)  \right) ^{2}x \left( s \right) -2\,\sqrt {2} \left( y
 \left( s \right)  \right) ^{2}{\it c1} \right) Q \left( s \right) }{{
{\it c0}}^{2}{a}^{3}}}+\]\[+{\frac { \left( -\sqrt {2}y \left( s \right) a{ \it
c0}\,{\it c2}+\sqrt {2} \left( y \left( s \right)  \right) ^{2}a{ \it c0} \right)
{\frac {d}{ds}}Q \left( s \right) }{{{\it c0}}^{2}{a}^ {3}}}+{\frac { \left( \sqrt
{2}y \left( s \right) a{\it c0}\,x \left( s \right) -\sqrt {2}y \left( s \right)
a{\it c0}\,{\it c1} \right) { \frac {d}{ds}}P \left( s \right) }{{{\it
c0}}^{2}{a}^{3}}}+\]\[+{\frac {
 \left( -2\,{a}^{2}{\it c0}\,x \left( s \right) +2\,{a}^{2}{\it c0}\,{
\it c1} \right) {\frac {d}{ds}}V \left( s \right) }{{{\it c0}}^{2}{a}^ {3}}}+{\frac
{ \left( -2\,a{\it c2}\,y \left( s \right) +2\,a \left( y
 \left( s \right)  \right) ^{2} \right) V \left( s \right) }{{{\it c0}
}^{2}{a}^{3}}}.
\end{equation}

   The relation (\ref{dryuma:eq23}) in this case takes a form
\begin{equation}\label{dryuma:eq27}
-1/2\,{\frac { \left( -2\,\alpha\,{\it c3}\,a+{\it c0}\,a \right) s}{{ \it
c0}\,a}}-1/2\,{\frac { \left( {\it c2}\,\sqrt {2}a-2\,y \left( s
 \right) \sqrt {2}a \right) V \left( s \right) }{{\it c0}\,a}}-1/2\,{
\frac { \left( 2\,{\it c1}\,y \left( s \right) -2\,x \left( s \right) y \left( s
\right)  \right) Q \left( s \right) }{{\it c0}\,a}}-\]\[-1/2\,{ \frac { \left( 2\,
\left( y \left( s \right)  \right) ^{2}-2\,{\it c2} \,y \left( s \right)  \right) P
\left( s \right) }{{\it c0}\,a}}-\nu=0
\end{equation}
and the system from three equations (\ref{dryuma:eq24}-\ref{dryuma:eq26}) can be
reduced to the system of two coupled equations.

   As example the substitution of the expression
\[
Q \left( s \right) =1/2\,{\frac { \left( -2\,\alpha\,{\it c3}\,a+{\it c0}\,a \right)
s}{y \left( s \right)  \left( x \left( s \right) -{\it c1} \right) }}+1/2\,{\frac {
\left( {\it c2}\,\sqrt {2}a-2\,y \left( s
 \right) \sqrt {2}a \right) V \left( s \right) }{y \left( s \right)
 \left( x \left( s \right) -{\it c1} \right) }}+1/2\,{\frac { \left( 2
\, \left( y \left( s \right)  \right) ^{2}-2\,{\it c2}\,y \left( s
 \right)  \right) P \left( s \right) }{y \left( s \right)  \left( x
 \left( s \right) -{\it c1} \right) }}+\]\[+{\frac {\nu\,{\it c0}\,a}{y
 \left( s \right)  \left( x \left( s \right) -{\it c1} \right) }}
\]
into the equation for $Q(s)$ give us the identity and in result our system takes the
form
\begin{equation}\label{dryuma:eq28}
{\frac {d^{2}}{d{s}^{2}}}P \left( s \right) =-{\frac { \left( -{\it c2 }+y \left( s
\right)  \right) {\it c2}\,{\frac {d}{ds}}P \left( s
 \right) }{{\it c0}\,a \left( x \left( s \right) -{\it c1} \right) }}-
1/2\,{\frac { \left( 2\, \left( x \left( s \right)  \right) ^{2}-4\,x
 \left( s \right) {\it c1}-2\,{\it c2}\,y \left( s \right) +{{\it c2}}
^{2}+2\,{{\it c1}}^{2} \right) \sqrt {2}{\frac {d}{ds}}V \left( s
 \right) }{{\it c0}\, \left( x \left( s \right) -{\it c1} \right) y
 \left( s \right) }}-\]\[-{\frac {{\it c2}\,y \left( s \right)  \left(
 \left( y \left( s \right)  \right) ^{2}+{{\it c2}}^{2}-2\,{\it c2}\,y
 \left( s \right) +{{\it c1}}^{2}-2\,x \left( s \right) {\it c1}+
 \left( x \left( s \right)  \right) ^{2} \right) P \left( s \right) }{
{a}^{2}{{\it c0}}^{2} \left(  \left( x \left( s \right)  \right) ^{2}- 2\,x \left( s
\right) {\it c1}+{{\it c1}}^{2} \right) }}+L(s)V \left( s
 \right) -\]\[-1/2\,{\frac { \left( -2\,\alpha\,{\it c3}\,a{\it c2}+{\it c0
}\,a{\it c2} \right) sy \left( s \right) }{{a}^{2}{{\it c0}}^{2}
 \left(  \left( x \left( s \right)  \right) ^{2}-2\,x \left( s
 \right) {\it c1}+{{\it c1}}^{2} \right) }}-1/2\,{\frac { \left( -{
\it c0}\,a{{\it c2}}^{2}+2\,\alpha\,{\it c3}\,a{{\it c2}}^{2} \right)
s}{{a}^{2}{{\it c0}}^{2} \left(  \left( x \left( s \right)  \right) ^{ 2}-2\,x
\left( s \right) {\it c1}+{{\it c1}}^{2} \right) }}-\]\[-1/2\,{ \frac { \left(
\left( 2\,\alpha\,{\it c3}\,a{\it c2}-{\it c0}\,a{\it c2} \right)  \left( x \left( s
\right)  \right) ^{2}+ \left( 2\,{\it c1}\,{\it c0}\,a{\it c2}-4\,\alpha\,{\it
c3}\,{\it c1}\,a{\it c2}
 \right) x \left( s \right) +2\,\alpha\,{\it c3}\,{{\it c1}}^{2}a{\it
c2}-{{\it c1}}^{2}{\it c0}\,a{\it c2} \right) s}{{a}^{2}{{\it
c0}}^{2} y \left( s \right)  \left(  \left( x \left( s \right)
\right) ^{2}-2 \,x \left( s \right) {\it c1}+{{\it c1}}^{2}
\right) }},
 \end{equation}
 where $$ L \left( s \right) ={\frac {
\left( y \left( s \right)  \right) ^{2} \sqrt {2}{\it c2}}{a{{\it
c0}}^{2} \left(  \left( x \left( s \right)
 \right) ^{2}-2\,x \left( s \right) {\it c1}+{{\it c1}}^{2} \right) }}
+1/2\,{\frac {\sqrt {2} \left( 2\, \left( x \left( s \right)  \right) ^{2}-3\,{{\it
c2}}^{2}-4\,x \left( s \right) {\it c1}+2\,{{\it c1}}^{2 } \right) y \left( s
\right) }{a{{\it c0}}^{2} \left(  \left( x
 \left( s \right)  \right) ^{2}-2\,x \left( s \right) {\it c1}+{{\it
c1}}^{2} \right) }}+\]\[+1/2\,{\frac {\sqrt {2} \left( -2\,{\it c2}\,
 \left( x \left( s \right)  \right) ^{2}-2\,{{\it c1}}^{2}{\it c2}+{{
\it c2}}^{3}+4\,{\it c2}\,{\it c1}\,x \left( s \right)  \right) }{a{{ \it c0}}^{2}
\left(  \left( x \left( s \right)  \right) ^{2}-2\,x
 \left( s \right) {\it c1}+{{\it c1}}^{2} \right) }}+\]\[+1/2\,{\frac {
\sqrt {2} \left( 2\, \left( x \left( s \right)  \right) ^{4}-8\,
 \left( x \left( s \right)  \right) ^{3}{\it c1}+ \left( {{\it c2}}^{2
}+12\,{{\it c1}}^{2} \right)  \left( x \left( s \right)  \right) ^{2}+
 \left( -2\,{{\it c2}}^{2}{\it c1}-8\,{{\it c1}}^{3} \right) x \left(
s \right) +{{\it c1}}^{2}{{\it c2}}^{2}+2\,{{\it c1}}^{4} \right) }{a{ {\it
c0}}^{2}y \left( s \right)  \left(  \left( x \left( s \right)
 \right) ^{2}-2\,x \left( s \right) {\it c1}+{{\it c1}}^{2} \right) }},
$$
and
\\[1mm]
\begin{equation}\label{dryuma:eq29}
{\frac {d^{2}}{d{s}^{2}}}V \left( s \right) =-{\frac { \left( 2\,
 \left( x \left( s \right)  \right) ^{2}-4\,x \left( s \right) {\it c1
}-3\,{\it c2}\,y \left( s \right) +2\, \left( y \left( s \right)
 \right) ^{2}+2\,{{\it c1}}^{2}+{{\it c2}}^{2} \right) {\frac {d}{ds}}
V \left( s \right) }{a{\it c0}\, \left( x \left( s \right) -{\it c1}
 \right) }}+\]\[+{\frac { \left(  \left( y \left( s \right)  \right) ^{2}+{
{\it c2}}^{2}-2\,{\it c2}\,y \left( s \right) +{{\it c1}}^{2}-2\,x
 \left( s \right) {\it c1}+ \left( x \left( s \right)  \right) ^{2}
 \right) \sqrt {2}y \left( s \right) {\frac {d}{ds}}P \left( s
 \right) }{{\it c0}\,{a}^{2} \left( x \left( s \right) -{\it c1}
 \right) }}+M(s)V \left( s \right) +N(s)P \left( s \right) +\]\[+1/2\,{\frac {
\sqrt {2} \left( {\it c0}-2\,\alpha\,{\it c3} \right) s \left( y
 \left( s \right)  \right) ^{3}}{{a}^{2}{{\it c0}}^{2} \left( x
 \left( s \right) -{{\it c1}}^{2} \right) }}+1/2\,{\frac {\sqrt {2}
 \left( -2\,{\it c0}\,{\it c2}+4\,\alpha\,{\it c3}\,{\it c2} \right) s
 \left( y \left( s \right)  \right) ^{2}}{{a}^{2}{{\it c0}}^{2}
 \left( x \left( s \right) -{{\it c1}}^{2} \right) }}\!+\!\]\[\!+\!1/2\,{\frac {
\sqrt {2} \left( -2\,\alpha\,{\it c3}\,{{\it c2}}^{2}\!+\!{{\it c2}}^{2}{ \it
c0}\!+\!{{\it c1}}^{2}{\it c0}\!+\! \left( x \left( s \right)  \right) ^{2 }{\it
c0}\!-\!2\,x \left( s \right) {\it c0}\,{\it c1}\!-\!2\,\alpha\,{\it c3} \,{{\it
c1}}^{2}\!+\!4\,\alpha\,{\it c3}\,x \left( s \right) {\it c1}\!-\!2\, \alpha\,{\it
c3}\, \left( x \left( s \right) \right) ^{2} \right) sy
 \left( s \right) }{{a}^{2}{{\it c0}}^{2} \left( x \left( s \right) -{
{\it c1}}^{2} \right) }},
\end{equation}
where
$$
M \left( s \right) =-2\,{\frac { \left( y \left( s \right)  \right) ^{
4}}{{a}^{2}{{\it c0}}^{2} \left(  \left( x \left( s \right)  \right) ^ {2}-2\,x
\left( s \right) {\it c1}+{{\it c1}}^{2} \right) }}+5\,{ \frac { \left( y \left( s
\right)  \right) ^{3}{\it c2}}{{a}^{2}{{\it c0}}^{2} \left(  \left( x \left( s
\right)  \right) ^{2}-2\,x \left( s
 \right) {\it c1}+{{\it c1}}^{2} \right) }}-\]\[-{\frac { \left( -4\,x
 \left( s \right) {\it c1}+2\, \left( x \left( s \right)  \right) ^{2}
+4\,{{\it c2}}^{2}+2\,{{\it c1}}^{2} \right)  \left( y \left( s
 \right)  \right) ^{2}}{{a}^{2}{{\it c0}}^{2} \left(  \left( x \left(
s \right)  \right) ^{2}-2\,x \left( s \right) {\it c1}+{{\it c1}}^{2}
 \right) }}-{\frac { \left( -{\it c2}\, \left( x \left( s \right)
 \right) ^{2}-{{\it c1}}^{2}{\it c2}+2\,{\it c2}\,{\it c1}\,x \left( s
 \right) -{{\it c2}}^{3} \right) y \left( s \right) }{{a}^{2}{{\it c0}
}^{2} \left(  \left( x \left( s \right)  \right) ^{2}-2\,x \left( s
 \right) {\it c1}+{{\it c1}}^{2} \right) }},
$$
\\[1mm]
$$
N \left( s \right) ={\frac {\sqrt {2} \left( y \left( s \right)
 \right) ^{5}}{{a}^{3}{{\it c0}}^{2} \left(  \left( x \left( s
 \right)  \right) ^{2}-2\,x \left( s \right) {\it c1}+{{\it c1}}^{2}
 \right) }}-3\,{\frac {\sqrt {2} \left( y \left( s \right)  \right) ^{
4}{\it c2}}{{a}^{3}{{\it c0}}^{2} \left(  \left( x \left( s \right)
 \right) ^{2}-2\,x \left( s \right) {\it c1}+{{\it c1}}^{2} \right) }}
+\]\[+{\frac {\sqrt {2} \left( 2\, \left( x \left( s \right)  \right) ^{2}+ 3\,{{\it
c2}}^{2}+2\,{{\it c1}}^{2}-4\,x \left( s \right) {\it c1}
 \right)  \left( y \left( s \right)  \right) ^{3}}{{a}^{3}{{\it c0}}^{
2} \left(  \left( x \left( s \right)  \right) ^{2}-2\,x \left( s
 \right) {\it c1}+{{\it c1}}^{2} \right) }}+{\frac {\sqrt {2} \left( -
3\,{\it c2}\, \left( x \left( s \right)  \right) ^{2}-{{\it c2}}^{3}-3 \,{{\it
c1}}^{2}{\it c2}+6\,{\it c2}\,{\it c1}\,x \left( s \right)
 \right)  \left( y \left( s \right)  \right) ^{2}}{{a}^{3}{{\it c0}}^{
2} \left(  \left( x \left( s \right)  \right) ^{2}-2\,x \left( s
 \right) {\it c1}+{{\it c1}}^{2} \right) }}+\]\[+{\frac {\sqrt {2} \left(
 \left( x \left( s \right)  \right) ^{4}-4\, \left( x \left( s
 \right)  \right) ^{3}{\it c1}+ \left( 6\,{{\it c1}}^{2}+{{\it c2}}^{2
} \right)  \left( x \left( s \right)  \right) ^{2}+ \left( -2\,{{\it c2}}^{2}{\it
c1}-4\,{{\it c1}}^{3} \right) x \left( s \right) +{{\it c1}}^{4}+{{\it c1}}^{2}{{\it
c2}}^{2} \right) y \left( s \right) }{{a} ^{3}{{\it c0}}^{2} \left(  \left( x \left(
s \right)  \right) ^{2}-2\, x \left( s \right) {\it c1}+{{\it c1}}^{2} \right) }}.
$$

     The expressions for functions $x(s)$ and $y(s)$ are dependent from the choice of parameters
     and can be defined from the equations (\ref{dryuma:eq16}).

     The integration of the equations (\ref{dryuma:eq28}-\ref{dryuma:eq29}) for
     the additional coordinates $P(s)$, $Q(s)$ is reduced to
    investigation of the $2\times 2$ system of second order ODE's with variable coefficients.

     Remark that the matrix $E$ and its properties play important role in analysis of such type
      of the system of equations.

    In result we get the correspondence between the geodesic in the $x,y,x,t$-space and the
    geodesic in the space with local coordinates $P,Q,U,V$ (partner space).

     The studying of such type of correspondence may be useful  from various point of view.

\section{Translation surfaces of the G\"odel spaces}

  Now we discuss the properties of translation surfaces of the G\"odel spaces.

   According with definition (\cite{dryuma15:eisen}) translation surfaces in arbitrary Riemannian space
   are defined by the systems of equations for local
   coordinates $x^{i}(u,v)$ of the space
  \begin{equation}\label{dryuma:eq30}
\frac{\partial x^{i}(u,v)}{\partial u \partial v}+\Gamma^{i}_{j
k}(x^{m})\frac{\partial x^{j}(u,v)}{\partial u }\frac{\partial
x^{k}(u,v)}{\partial v}=0,
\end{equation}
where $\Gamma^{i}_{jk}$ are the Christoffel coefficients.

     In the case of the G\"odel  metric (\ref{dryuma:eq6}) we get the equations
\begin{equation}\label{dryuma:eq31}
{\frac {\partial ^{2}}{\partial u\partial v}}x \left( u,v \right)
+1/2 \,{\frac {\sqrt {2} \left(  \left( {\frac {\partial
}{\partial u}}y
 \left( u,v \right)  \right) {\frac {\partial }{\partial v}}t \left( u
,v \right) + \left( {\frac {\partial }{\partial u}}t \left( u,v
 \right)  \right) {\frac {\partial }{\partial v}}y \left( u,v \right)
 \right) }{a}}=0,
\end{equation}
\\[1mm]
\begin{equation}\label{dryuma:eq32}
{\frac {\partial ^{2}}{\partial u\partial v}}y \left( u,v \right)
-1/2 \,{\frac {\sqrt {2} \left( {\frac {\partial }{\partial v}}x
\left( u,v
 \right)  \right) {\frac {\partial }{\partial u}}t \left( u,v \right)
}{a}}-1/2\,{\frac {\sqrt {2} \left( {\frac {\partial }{\partial
v}}t
 \left( u,v \right)  \right) {\frac {\partial }{\partial u}}x \left( u
,v \right) }{a}}-\]\[-{\frac { \left( {\frac {\partial }{\partial
u}}x
 \left( u,v \right)  \right) {\frac {\partial }{\partial v}}x \left( u
,v \right) + \left( {\frac {\partial }{\partial u}}y \left( u,v
 \right)  \right) {\frac {\partial }{\partial v}}y \left( u,v \right)
}{y \left( u,v \right) }}=0,
\end{equation}
\\[1mm]
\begin{equation}\label{dryuma:eq33}
{\frac {\partial ^{2}}{\partial u\partial v}}t \left( u,v \right)
-{ \frac { \left( {\frac {\partial }{\partial u}}y \left( u,v
\right)
 \right) {\frac {\partial }{\partial v}}t \left( u,v \right) + \left(
{\frac {\partial }{\partial u}}t \left( u,v \right)  \right)
{\frac {
\partial }{\partial v}}y \left( u,v \right) }{y \left( u,v \right) }}-\]\[-
{\frac {1/2\,\sqrt {2}a \left( {\frac {\partial }{\partial u}}x
 \left( u,v \right)  \right) {\frac {\partial }{\partial v}}y \left( u
,v \right) +1/2\,\sqrt {2}a \left( {\frac {\partial }{\partial
u}}y
 \left( u,v \right)  \right) {\frac {\partial }{\partial v}}x \left( u
,v \right) }{ \left( y \left( u,v \right)  \right) ^{2}}}=0
\end{equation}
\\[1mm]
\begin{equation}\label{dryuma:eq34}
{\frac {\partial ^{2}}{\partial u\partial v}}z \left( u,v
\right)=0
\end{equation}

    Full integration of this nonlinear system of equations is a difficult
    problem.

    To cite one example.

    With this aim we present our system of equations in a new
    coordinates $u = r + s, v = r - s$.

    It takes the form
$$ 2\,\left (1/4\,{\frac {\partial ^{2}}{\partial
{r}^{2}}}x(r,s)\!-\!1/4\,{ \frac {\partial ^{2}}{\partial
{s}^{2}}}x(r,s)\right )a\!+\!\sqrt {2} \left (1/2\,{\frac
{\partial }{\partial r}}y(r,s)\!+\!1/2\,{\frac {
\partial }{\partial s}}y(r,s)\right )\left (1/2\,{\frac {\partial }{
\partial r}}t(r,s)\!-\!1/2\,{\frac {\partial }{\partial s}}t(r,s)\right
)+\]\[\!+\! \sqrt {2}\left (1/2\,{\frac {\partial }{\partial
r}}t(r,s)\!+\!1/2\,{ \frac {\partial }{\partial s}}t(r,s)\right
)\left (1/2\,{\frac {
\partial }{\partial r}}y(r,s)\!-\!1/2\,{\frac {\partial }{\partial s}}y(r,
s)\right )=0,
$$
\\[1mm]
 $$ -2\,\left (1/4\,{\frac {\partial
^{2}}{\partial {r}^{2}}}y(r,s)-1/4\,{ \frac {\partial
^{2}}{\partial {s}^{2}}}y(r,s)\right )y(r,s) a+\]\[+2\, \left
(1/2\,{\frac {\partial }{\partial r}}x(r,s)+1/2\,{\frac {
\partial }{\partial s}}x(r,s)\right )\left (1/2\,{\frac {\partial }{
\partial r}}x(r,s)-1/2\,{\frac {\partial }{\partial s}}x(r,s)\right )a
\!+\!\]\[+\sqrt {2}\left (1/2\,{\frac {\partial }{\partial
r}}x(r,s)+1/2\,{ \frac {\partial }{\partial s}}x(r,s)\right )\left
(1/2\,{\frac {
\partial }{\partial r}}t(r,s)-1/2\,{\frac {\partial }{\partial s}}t(r,
s)\right )y(r,s)+\]\[+2\,\left (1/2\,{\frac {\partial }{\partial
r}}y(r,s)+1/2\,{\frac {\partial }{\partial s}}y(r,s)\right )\left
(1/2\,{\frac {
\partial }{\partial r}}y(r,s)-1/2\,{\frac {\partial }{\partial s}}y(r,
s)\right )a+\]\[+\sqrt {2}\left (1/2\,{\frac {\partial }{\partial
r}}t(r,s)+1/2\,{\frac {\partial }{\partial s}}t(r,s)\right )\left
(1/2\,{\frac {\partial }{\partial r}}x(r,s)-1/2\,{\frac {\partial
}{\partial s}}x(r ,s)\right )y(r,s)=0,
 $$
\\[1mm]
$$ -2\,\left (1/4\,{\frac {\partial ^{2}}{\partial
{r}^{2}}}t(r,s)-1/4\,{ \frac {\partial ^{2}}{\partial
{s}^{2}}}t(r,s)\right )\left (y(r,s) \right )^{2}+\]\[+\sqrt
{2}a\left (1/2\,{\frac {\partial }{\partial r}}x(r, s)+1/2\,{\frac
{\partial }{\partial s}}x(r,s)\right )\left (1/2\,{ \frac
{\partial }{\partial r}}y(r,s)-1/2\,{\frac {\partial }{\partial
s}}y(r,s)\right )+\]\[+\sqrt {2}a\left (1/2\,{\frac {\partial
}{\partial r} }y(r,s)+1/2\,{\frac {\partial }{\partial
s}}y(r,s)\right )\left (1/2\, {\frac {\partial }{\partial
r}}x(r,s)-1/2\,{\frac {\partial }{
\partial s}}x(r,s)\right )+\]\[+2\,\left (1/2\,{\frac {\partial }{\partial
r}}y(r,s)+1/2\,{\frac {\partial }{\partial s}}y(r,s)\right )\left
(1/2 \,{\frac {\partial }{\partial r}}t(r,s)-1/2\,{\frac {\partial
}{
\partial s}}t(r,s)\right )y(r,s)+\]\[+2\,\left (1/2\,{\frac {\partial }{
\partial r}}t(r,s)+1/2\,{\frac {\partial }{\partial s}}t(r,s)\right )
\left (1/2\,{\frac {\partial }{\partial r}}y(r,s)-1/2\,{\frac {
\partial }{\partial s}}y(r,s)\right )y(r,s)=0.
$$

   The solution of this system of equations we shall seek in form
$$
 y(r,s)=B(r),\quad t(r,s)=C(r)-s,\quad x(r,s)=s+A(r),
$$ where $B(r),C(r),A(r)$ are  some unknown functions.

   In result our system takes the form
$$
\left ({\frac {d^{2}}{d{r}^{2}}}C(r)\right )\left (B(r)\right
)^{2}- \sqrt {2}a\left ({\frac {d}{dr}}B(r)\right ){\frac
{d}{dr}}A(r)-2\, \left ({\frac {d}{dr}}B(r)\right )B(r){\frac
{d}{dr}}C(r)=0,
 $$
$$ \left ({\frac {d^{2}}{d{r}^{2}}}B(r)\right )B(r)a\!-\!a\left
({\frac {d}{d r}}A(r)\right )^{2}\!+\!a-\sqrt {2}B(r)\left ({\frac
{d}{dr}}A(r)\right ){ \frac {d}{dr}}C(r)\!-\!\sqrt
{2}B(r)\!-\!\left ({\frac {d}{dr}}B(r)\right )^{2 }a=0, $$
 $$ \left ({\frac
{d^{2}}{d{r}^{2}}}A(r)\right )a+\sqrt {2}\left ({\frac {
d}{dr}}B(r)\right ){\frac {d}{dr}}C(r)=0 .
$$

    Using the first integral
$$ {\frac {d}{dr}}C(r)=-{\frac {\sqrt {2}a{\frac
{d}{dr}}A(r)}{B(r)}}+ \alpha, $$ the system can be written in form
$$
 \left ({\frac {d^{2}}{d{r}^{2}}}B(r)\right )B(r)a+a\left
({\frac {d}{d r}}A(r)\right )^{2}+a-\sqrt {2}\left ({\frac
{d}{dr}}A(r)\right ) \alpha\,B(r)-\sqrt {2}B(r)-\left ({\frac
{d}{dr}}B(r)\right )^{2}a=0,
 $$
$$\left ({\frac {d^{2}}{d{r}^{2}}}A(r)\right )aB(r)-2\,\left
({\frac {d} {dr}}B(r)\right )a{\frac {d}{dr}}A(r)+\sqrt {2}\left
({\frac {d}{dr}}B (r)\right )\alpha\,B(r)=0.
 $$

   Integration of the last equation give us the expression for
   the function $A(r)$
$$ A(r)=\int \!{\frac {B(r)\left (\sqrt {2}\alpha+B(r){\it C_1}\,a
\right )}{a}}{dr}+{\it C_2}
 $$
with parameters $C_1,C_2$ and $\alpha $.

   After substitution the expression for $A(r)$ to the first equation  we get
   the equation
$$ \left ({\frac {d^{2}}{d{r}^{2}}}B(r)\right )B(r)a+\left
(B(r)\right )^ {3}\sqrt {2}\alpha\,{\it C_1}+\left (B(r)\right
)^{4}{{\it C_1}}^{2} a+a-\sqrt {2}B(r)-\left ({\frac
{d}{dr}}B(r)\right )^{2}a=0 $$
 for the function $B(r)$.

Remark that this equation can be written in form
 $$
 {\frac {d^{2}}{d{r}^{2}}}E(r)={\frac {\sqrt {2}{e^{-E(r)}}}{a}}-{{\it
C_1}}^{2}{e^{2 E(r)}}-{\frac {\sqrt {2}\alpha\,{\it
C_1}\,{e^{E(r)} }}{a}}-{e^{-2 E(r)}}
 $$
after the change of variable $$ B(r)=\exp(E(r)).
$$

    With the help of its solutions  the examples of  the translation surfaces
    of the G\"odel space (\ref{dryuma:eq6}) can be constructed.

    They can be presented  in form
$$
 t+x=A(r)+C(r),\quad y=B(r),
$$ or
$$
t(x,y)=x+\phi(y)
 $$
 with some function $\phi(y)$.

  Detail consideration of the properties of this type of translation surfaces, their intrinsic geometry
  and characteristic lines will be done in a following publications of author.

   Remark that with the help of the translation  surfaces can be investigated the
   properties of closed trajectories of the G\"odel space.

   Let us consider the eight-dimensional extension of the G\"odel space
   with the metric (\ref{dryuma:eq17}).

   Translation surfaces in this case are determined by the equations
(\ref{dryuma:eq28})-(\ref{dryuma:eq31}) for coordinates $x,y,z,t$
and by the linear system of equations on coordinates $P,Q,U,V$
\begin{equation} \label{dryuma:eq35}
{\frac {\partial ^{2}}{\partial u\partial v}}P \left( u,v \right)
-\]\[\!-\!1/2 \,{\frac { \left(\!-\!2\,\sqrt {2}{a}^{2} \left(
{\frac {\partial }{
\partial u}}x   \right) {\frac {\partial }{\partial
v}}x\!-\!2\, \left( {\frac {\partial }{\partial u}}x
   \right)  \left( {\frac {\partial }{\partial v}}t
   \right) ya\!+\!2\,\sqrt {2}{a}^{2} \left( {\frac {
\partial }{\partial u}}y   \right) {\frac {\partial
}{\partial v}}y \!-\!2\, \left( {\frac {\partial }{
\partial u}}t   \right)  \left( {\frac {\partial }{
\partial v}}x   \right) y a \right) V \left( u,v
 \right) }{{y}^{3}a}}\!-\]\[-\!1/2\,{\frac { \left( -2\, \left( {\frac {
\partial }{\partial u}}x \left( u,v \right)  \right) {y}^{2}a-\sqrt {2
} \left( {\frac {\partial }{\partial u}}t \left( u,v \right)
\right) {y}^{3} \right) {\frac {\partial }{\partial v}}Q \left(
u,v \right) }{ {y}^{3}a}}-\]\[-1/2\,{\frac { \left( -\sqrt {2}
\left( {\frac {\partial }{
\partial v}}t \left( u,v \right)  \right) {y}^{3}-2\, \left( {\frac {
\partial }{\partial v}}x \left( u,v \right)  \right) {y}^{2}a \right)
{\frac {\partial }{\partial u}}Q \left( u,v \right)
}{{y}^{3}a}}+\]\[+1/2\, {\frac {\sqrt {2}a \left( {\frac {\partial
}{\partial u}}y \left( u,v
 \right)  \right) {\frac {\partial }{\partial v}}V \left( u,v \right)
}{{y}^{2}}}+1/2\,{\frac {\sqrt {2}a \left( {\frac {\partial }{
\partial v}}y \left( u,v \right)  \right) {\frac {\partial }{\partial
u}}V \left( u,v \right) }{{y}^{2}}} =0,
\end{equation}
\\[1mm]
\begin{equation} \label{dryuma:eq36}
{\frac {\partial ^{2}}{\partial u\partial v}}Q \left( u,v \right)
+\]\[+1/2 \,{\frac { \left( 6\, \left( {\frac {\partial }{\partial
u}}x   \right) \left( {\frac {\partial }{\partial v}}x  \right)
ya+2\,\sqrt {2} \left( {\frac {\partial }{\partial u}}x   \right)
\left( {\frac {\partial }{\partial v }}t  \right) {y}^{2}+2\,\sqrt
{2} \left( {\frac {
\partial }{\partial u}}t   \right)  \left( {\frac {
\partial }{\partial v}}x   \right) {y}^{2}+2\,
 \left( {\frac {\partial }{\partial u}}y   \right)
 \left( {\frac {\partial }{\partial v}}y   \right) y
a \right) Q \left( u,v \right) }{{y}^{3}a}}+\]\[+1/2\,{\frac {
\left( -2\,a
 \left( {\frac {\partial }{\partial u}}x   \right)
 \left( {\frac {\partial }{\partial v}}y   \right) y
-2\, \left( {\frac {\partial }{\partial u}}y
 \right)  \left( {\frac {\partial }{\partial v}}t
 \right) {y}^{2}\sqrt {2}-2\, \left( {\frac {\partial }{\partial u}}t
  \right)  \left( {\frac {\partial }{\partial v}}y
   \right) {y}^{2}\sqrt {2}-2\,a \left( {\frac {
\partial }{\partial u}}y   \right)  \left( {\frac {
\partial }{\partial v}}x \right) y \right) P
 \left( u,v \right) }{{y}^{3}a}}+\]\[+1/2\,{\frac { \left( 2\,\sqrt {2}{a}^
{2} \left( {\frac {\partial }{\partial u}}x
 \right) {\frac {\partial }{\partial v}}y  +2\,
 \left( {\frac {\partial }{\partial u}}y  \right)
 \left( {\frac {\partial }{\partial v}}t   \right) y
a+2\,\sqrt {2}{a}^{2} \left( {\frac {\partial }{\partial u}}y
\right) {\frac {\partial }{\partial v}}x +2\, \left( {\frac
{\partial }{\partial u}}t\right) \left( {\frac {\partial
}{\partial v}}y  \right) ya \right) V \left( u,v \right)
}{{y}^{3}a}}+\]\[+{\frac { \left( {\frac {\partial }{\partial u}}y
\left( u,v \right) \right) {\frac {\partial }{\partial v}}Q \left(
u,v \right) }{y}}+{\frac {
 \left( {\frac {\partial }{\partial v}}y \left( u,v \right)  \right) {
\frac {\partial }{\partial u}}Q \left( u,v \right)
}{y}}+1/2\,{\frac {
 \left( 2\, \left( {\frac {\partial }{\partial u}}t \left( u,v
 \right)  \right) {y}^{2}a+\sqrt {2}{a}^{2} \left( {\frac {\partial }{
\partial u}}x \left( u,v \right)  \right) y \right) {\frac {\partial }
{\partial v}}V \left( u,v \right) }{{y}^{3}a}}+\]\[+1/2\,{\frac {
\left( \sqrt {2}{a}^{2} \left( {\frac {\partial }{\partial v}}x
\left( u,v
 \right)  \right) y+2\, \left( {\frac {\partial }{\partial v}}t
 \left( u,v \right)  \right) {y}^{2}a \right) {\frac {\partial }{
\partial u}}V \left( u,v \right) }{{y}^{3}a}}+\]\[+1/2\,{\frac {-\sqrt {2}
 \left( {\frac {\partial }{\partial u}}t \left( u,v \right)  \right)
 \left( {\frac {\partial }{\partial v}}P \left( u,v \right)  \right) {
y}^{3}-\sqrt {2} \left( {\frac {\partial }{\partial u}}P \left(
u,v
 \right)  \right)  \left( {\frac {\partial }{\partial v}}t \left( u,v
 \right)  \right) {y}^{3}}{{y}^{3}a}}=0,
\end{equation}
\\[1mm]
\begin{equation}\label{dryuma:eq37}
{\frac {\partial ^{2}}{\partial u\partial v}}V \left( u,v \right)
+\]\[+1/2 \,{\frac { \left( -2\, \left( {\frac {\partial
}{\partial u}}y \left( u,v \right)  \right) \left( {\frac
{\partial }{\partial v}}t \left( u ,v \right)  \right) {y}^{2}-2\,
\left( {\frac {\partial }{\partial u}} t \left( u,v \right)
\right)  \left( {\frac {\partial }{\partial v}}y
 \left( u,v \right)  \right) {y}^{2} \right) Q \left( u,v \right) }{{a
}^{2}{y}^{2}}}+\]\[+1/2\,{\frac { \left( -2\, \left( {\frac
{\partial }{
\partial u}}x \right)  \left( {\frac {\partial }{
\partial v}}t   \right) {y}^{2}-2\, \left( {\frac {
\partial }{\partial u}}x   \right)  \left( {\frac {
\partial }{\partial v}}x   \right) a\sqrt {2}y-2\,
\sqrt {2} \left( {\frac {\partial }{\partial u}}y
 \right)  \left( {\frac {\partial }{\partial v}}y
 \right) ay-2\, \left( {\frac {\partial }{\partial u}}t  \right)
  \left( {\frac {\partial }{\partial v}}x  \right) {y}^{2} \right) P \left( u,v \right) }{{a}^{2}{y}^{2
}}}+\]\[+1/2\,{\frac { \left( 4\, \left( {\frac {\partial
}{\partial u}}x
   \right)  \left( {\frac {\partial }{\partial v}}x
   \right) {a}^{2}+2\,\sqrt {2} \left( {\frac {
\partial }{\partial u}}x  \right)  \left( {\frac {
\partial }{\partial v}}t  \right) ya+2\,\sqrt {2}
 \left( {\frac {\partial }{\partial u}}t   \right)
 \left( {\frac {\partial }{\partial v}}x   \right) y
a \right) V \left( u,v \right) }{{a}^{2}{y}^{2}}}+\]\[+1/2\,{\frac
{\sqrt { 2} \left( {\frac {\partial }{\partial u}}x \left( u,v
\right)
 \right) {\frac {\partial }{\partial v}}Q \left( u,v \right) }{a}}+\]\[+1/2
\,{\frac {\sqrt {2} \left( {\frac {\partial }{\partial u}}Q \left(
u,v
 \right)  \right) {\frac {\partial }{\partial v}}x \left( u,v \right)
}{a}}+{\frac { \left( {\frac {\partial }{\partial u}}y \left( u,v
 \right)  \right) {\frac {\partial }{\partial v}}V \left( u,v \right)
}{y}}+{\frac { \left( {\frac {\partial }{\partial u}}V \left( u,v
 \right)  \right) {\frac {\partial }{\partial v}}y \left( u,v \right)
}{y}}+\]\[+1/2\,{\frac {-\sqrt {2} \left( {\frac {\partial
}{\partial u}}P
 \left( u,v \right)  \right)  \left( {\frac {\partial }{\partial v}}y
 \left( u,v \right)  \right) a{y}^{2}-\sqrt {2} \left( {\frac {
\partial }{\partial u}}y \left( u,v \right)  \right)  \left( {\frac {
\partial }{\partial v}}P \left( u,v \right)  \right) a{y}^{2}}{{a}^{2}
{y}^{2}}}=0
\end{equation}
\begin{equation}\label{dryuma:eq38}
{\frac {\partial ^{2}}{\partial u\partial v}}U \left( u,v
\right)=0.
\end{equation}

    The system of linear equations  (\ref{dryuma:eq32})-(\ref{dryuma:eq35})
    is the matrix analog of the Laplace equation
\begin{equation}\label{dryuma:eq39}
\frac{\partial^2 \Psi(u,v)}{\partial u \partial
v}+A(u,v)\frac{\partial \Psi(u,v)}{\partial u
}+B(u,v)\frac{\partial \Psi(u,v)}{\partial v}+C(u,v)\Psi(u,v) =0,
\end{equation}
 where
\[
\Psi(u,v)=\left(\begin{array}{c} P(u,v)\\ Q(u,v)\\ U(u,v)\\ V(u,v)
\end{array}\right)
\]
is a vector-function, and $A(u,v),B(u,v),C(u,v)$ are the matrices
depending from the variables $(u,v)$.

    For integration of such type equation  can be used the matrix generalization of the
    Darboux Invariants ~(\cite {dryuma16:elianu}).
\begin{rem}

    We are reminded  basic facts on integration of the matrix Laplace-equation.

   The system (\ref{dryuma:eq39}) can be presented in forms
\[
(\partial_u+B)(\partial_v+A)\Psi-H\Psi=0,
\]
or
\[
(\partial_v+A)(\partial_u+B)\Psi-K\Psi=0,
\]
where
 \[
H=\frac{\partial A}{\partial u}+B A-C,\quad K=\frac{\partial
B}{\partial v}+A B-C,
\]
are the Darboux  invariants of the system.

   In the case $K=0$ or $H=0$ the system can be integrated.

   If the $K\neq0$ and $H\neq0$ the system may presented in similar form for the
   functions
\[
\Psi_1=\frac{\partial \Psi}{\partial y}+A \Psi.
\]
or
\[
\Psi_{-1}=\frac{\partial \Psi}{\partial x}+B \Psi.
\]

   In the first case one get
\[
\frac{\partial^2 \Psi_1(u,v)}{\partial u \partial
v}+A_1(u,v)\frac{\partial \Psi_1(u,v)}{\partial u
}+B_1(u,v)\frac{\partial \Psi_1(u,v)}{\partial
v}+C_1(u,v)\Psi_1(u,v) =0,
\]
where
\[
    A_1=H A H^{-1}-H_v H^{-1},\quad B_1=B,\quad C_1=B_v-H+(HA
    H^{-1}-H_v H^{-1})B.
\]

The invariants $H_1$ and $K_1$ for this equation are
\[
H_1=H-B_y+(HAH^{-1}-H_y H^{-1})_u+B(H A H^{-1}-H_v
H^{-1})-(HAH^{-1}-H_y H^{-1})B,\]\[ K_1=H.
\]

    In the case $H_1=0$ the system can be integrated.

   In the second case we get the equation for the function
   $\Psi_{-1}$
\[
\frac{\partial^2 \Psi_{-1}(u,v)}{\partial u \partial
v}+A_{-1}(u,v)\frac{\partial \Psi_{-1}(u,v)}{\partial u
}+B_{-1}(u,v)\frac{\partial \Psi_{-1}(u,v)}{\partial
v}+C_{-1}(u,v)\Psi_{-1}(u,v) =0,
\]
where
\[
    B_{-1}=K B K^{-1}-K_u K^{-1},\quad A_{-1}=A,\quad C_{-1}=A_u-K+(K B K^{-1}-K_u K^{-1})A.
\]

The invariants $H_{-1}$ and $K_{-1}$ for this equation are
\[
K_{-1}=K-A_u+(KBK^{-1}-K_u K^{-1})_v+A(K B K^{-1}-K_u
K^{-1})-(KBK^{-1}-K_u K^{-1})A,\]\[ H_{-1}=K
\]
and at the condition $K_{-1}=0$ the system is also integrable.

     To integrate the system of equations ~(\ref{dryuma:eq39})
     in explicit form  it is necessary to use the expressions for
     coordinates $x(u,v),y(u,v),z(u,v),t(u,v)$ of translation surfaces
     of the  basic space.

     The properties of the invariants $H$ and $K$
     also may be important for classifications of translation surfaces
      of the basic and extended G\"odel space.

\end{rem}

\section{On the spectrum of the G\"odel space-time metric}

    In this section the spectrum $\lambda$ of  de Rham operator
\[
\Delta=g^{ij}\nabla_i\nabla_j-Ricci,
\]
    defined on a four-dimensional riemannian manifold  and
    acting on 1-forms
$$
 \omega=A_i(x,y,z)dx^i=u(x,y,z,t)dx+v(x,y,z,t)dy
+p(x,y,z,t)dz +q(x,y,z,t)dt $$
 will be calculated.

    The problem is reduced to the solution of the system of equations
\begin{equation}\label{dryuma:eq40}
 g^{ij}\nabla_i\nabla_j A_k-R^l_{k}A_l-\mu^2 A_k=0,
\end{equation}
 where $\nabla_k$ is a symbol of covariant derivative and
the $R{^i_j}$ is the Ricci tensor of the metric $g^{ij}$ of the
G\"odel space-time.

   We use the  G\"odel space-time metric in form
   ~(\ref{dryuma:eq11}) and for simplicity sake the components of the 1-form $\omega$
    will be presented as
   $$
   A_k=[0,v(y,t),0,q(y,t)].
   $$

    As this takes place the system ~(\ref{dryuma:eq40})  looks as
\[ -{\frac {\partial }{\partial t}}v(y,t)+{\frac {\partial
}{\partial y}} q(y,t)=0,
 \]
\[
2\,\left ({\frac {\partial }{\partial y}}v(y,t)\right )y+\left ({
\frac {\partial ^{2}}{\partial {y}^{2}}}v(y,t)\right
){y}^{2}+{a}^{2}{ \frac {\partial ^{2}}{\partial
{t}^{2}}}v(y,t)-{\mu}^{2}v(y,t){a}^{2}=0,
\]
\[
2\,\left ({\frac {\partial }{\partial y}}q(y,t)\right )y-2\,\left
({ \frac {\partial }{\partial t}}v(y,t)\right )y+\left ({\frac
{\partial ^{2}}{\partial {y}^{2}}}q(y,t)\right ){y}^{2}+\left
({\frac {\partial ^{2}}{\partial {t}^{2}}}q(y,t)\right
){a}^{2}-{\mu}^{2}q(y,t){a}^{2}=0.
\]

    It is equivalent to the following non homogeneous equation
\begin{equation}\label{dryuma:eq41}
-\left ({\frac {\partial ^{2}}{\partial {y}^{2}}}\Phi(y,t)\right
){y}^ {2}-\left ({\frac {\partial ^{2}}{\partial
{t}^{2}}}\Phi(y,t)\right ){
a}^{2}+{\mu}^{2}\Phi(y,t){a}^{2}+\epsilon=0 , \end{equation}
 where
 \[
q(y,t)=\frac{ \partial \Phi(y,t)}{\partial t},v(y,t)=\frac
{\partial \Phi(y,t)}{\partial y}
 \]
and $\epsilon$  is parameter.

   The simplest solution of homogeneous equation can be presented in form

\begin{equation}\label{dryuma:eq42}
\Phi(y,t)={\it F_1}(y){\it F_2}(t),
\end{equation}
where
\begin{equation}\label{dryuma:eq43}
{\frac {d^{2}}{d{t}^{2}}}{\it F_2}(t)=-{\frac {{\it c}_{{1}}{\it
F_2}(t)}{{a}^{2}}}+{\mu}^{2}{\it F_2}(t),$$ $$ {\frac
{d^{2}}{d{y}^{2}}}{\it F_1}(y)={\frac {{\it c}_{{1}}{\it
F_1}(y)}{{y}^{2}}} ,
\end{equation}
and $c_1$ is the parameter.

   The second equation from ~(\ref{dryuma:eq43}) has the form
\[
{\frac {d^{2}}{d{y}^{2}}}{\it F_1}(y)-{\frac {{\it c}_{{1}}{\it
F_1}(y)}{{y}^{2}}}=0
\]
and its solutions are defined by the relations
 $$ \frac{F_1}{\sqrt{y}}=C_1 \cos(b \ln(y))+C_2 \sin(b\ln(y)),\quad
 b^2=-c_1-\frac{1}{4}>0,
 $$
$$  \frac{F_1}{\sqrt{y}}=C_1 x^b+C_2 x^{-b},\quad
b^2=\frac{1}{4}+c_1>0,$$ $$
\frac{F_1}{\sqrt{y}}=C_1+C_2\ln(y),\quad c_1=-\frac{1}{4},
 $$
which from the parameter $c_1$ are dependent.

    The solutions of the first equation of the system  ~(\ref{dryuma:eq43}) are
\[
{\it F_2}(t)={\it C_3}\,\sin(1/4\,{\frac {\sqrt {2}\sqrt {{\it c}_
{{1}}-8\,{\mu}^{2}{a}^{2}}t}{a}})+{\it C_4}\,\cos(1/4\,{\frac
{\sqrt {2}\sqrt {{\it c}_{{1}}-8\,{\mu}^{2}{a}^{2}}t}{a}}).
\]

    In result the general solution of the equation ~(\ref{dryuma:eq41}) can be constructed
    with the help of solutions $F_1(y)$ and $F_2(t)$.

    So depend upon the choice of $c_1$ the spectrum of manifold and the solutions of the equation
~(\ref{dryuma:eq41}) will be various.

     The problem of solutions of the system ~(\ref{dryuma:eq40}) in more general
     case of  the 1-form $
 \omega=A_i(x,y,z)dx^i$ requires more detail consideration.

\section{Spatial metric
of the four-dimensional G\"odel space-time}

    The spatial metric of any four-dimensional metric
\[
^{4}ds^2=g_{\alpha \beta}dx^{\alpha} dx^{\beta}+2g_{0\alpha} dx^{0}
dx^{\alpha}+g_{00}dx^{0} dx^{0}
\]
has the  form
\[
^{3}dl^2=\gamma_{\alpha \beta}dx^{\alpha}dx^{\beta},
\]
where
\[
\gamma_{\alpha \beta}=-g_{\alpha \beta}+\frac{g_{0\alpha}g_{0\beta}}{g_{00}}.
\]
 a three-dimensional tensor determining the properties of the
 space.

   In the case of the  G\"odel space-time the spatial three-dimensional metric
   has the form
\begin{equation}\label{dryuma:eq44}
-^{3}dl^2=\frac{a^2}{y^2}(dx^2+dy^2)+dz^2.
\end{equation}

     Three-dimensional space with the metric ~(\ref{dryuma:eq44}) belongs to the one of the
eight types of the W.Thurston geometries and has a diverse global properties.

   In particular it admits the surfaces bundle.

   As example we consider the translation surfaces of the space ~(\ref{dryuma:eq44}).

    They defined by the system of equations for coordinates
    $x(u,v),y(u,v)$ and $z(u,v)$
\begin{equation}\label{dryuma:eq45}
{\frac {\left ({\frac {\partial ^{2}}{\partial u\partial v}}x(u,v) \right
)y(u,v)-\left ({\frac {\partial }{\partial u}}x(u,v)\right ){ \frac {\partial
}{\partial v}}y(u,v)-\left ({\frac {\partial }{
\partial u}}y(u,v)\right ){\frac {\partial }{\partial v}}x(u,v)}{y(u,v
)}}=0, $$$$  {\frac {\left ({\frac {\partial ^{2}}{\partial
u\partial v}}y(u,v) \right )y(u,v)+\left ({\frac {\partial
}{\partial u}}x(u,v)\right ){ \frac {\partial }{\partial
v}}x(u,v)-\left ({\frac {\partial }{
\partial u}}y(u,v)\right ){\frac {\partial }{\partial v}}y(u,v)}{y(u,v
)}}=0, \end{equation}
\[
{\frac {\partial ^{2}}{\partial u\partial v}}z(u,v)=0 .
\]

    The simplest solutions of these equations are in form
\[
y(u,v)=1/2\,\left (1+\left (\left ({\frac {u}{v}}\right )^{{\it C_1}} \right
)^{2}\left ({e^{{\it C_2}\,{\it C_1}}}\right )^{-2}\right ){e ^{{\it C_2}\,{\it
C_1}}}\left (\left ({\frac {u}{v}}\right )^{{\it C_1}}\right )^{-1}{{\it C_1}}^{-1},
\]
\[
x(u,v)=\ln (u)+\ln (v),
\]
and
\[
z(u,v)=A(u)+B(v),
\]
where $A(v)$ and $B(v)$ are arbitrary functions and $C_1,C_2$ are the parameters.

 In particular case $C_1=1$, $C_2=0$ one get
\[
y(u,v)=1/2\,{\frac {{u}^{2}+{v}^{2}}{u v}},
\]
\[
x(u,v)=\ln (u v).
\]

   From here we find
\[
v=\sqrt {y{e^{x}}+{e^{x}}\sqrt {{y}^{2}-1}},
\]
\[
u={\frac {{e^{x}}}{\sqrt {y{e^{x}}+{e^{x}}\sqrt {{y}^{2}-1}}}}
\]
and
\[
z(x,y)=A(\sqrt {y{e^{x}}+{e^{x}}\sqrt {{y}^{2}-1}})+B({\frac {{e^{x}}}{\sqrt
{y{e^{x}}+{e^{x}}\sqrt {{y}^{2}-1}}}})
\]
with arbitrary functions $A(u), B(v)$.

   The properties of surfaces are dependent from the choice of the functions $A$ and $B$.

\begin{rem}

    From the system ~(\ref{dryuma:eq45}) we find the relations
$$ \left ({\frac {\partial }{\partial v}}x(u,v)\right )^{2}-{\frac
{{e^{2 \,z(u,v)}}}{{v}^{2}}}+\left ({\frac {\partial }{\partial
v}}z(u,v) \right )^{2}{e^{2\,z(u,v)}}=0,
 $$
$$
 \left ({\frac {\partial }{\partial u}}x(u,v)\right
)^{2}-{\frac {{e^{2 \,z(u,v)}}}{{u}^{2}}}+\left ({\frac {\partial
}{\partial u}}z(u,v) \right )^{2}{e^{2\,z(u,v)}}=0,
 $$
where $ z(u,v)=\ln(y(u,v))$.

   This fact allow us to get one equation on variable $y(u,v)$ only. $$
\left (-\sqrt {\left (y\right )^{2}\!-\!\left ({\frac {\partial }{
\partial v}}y\right )^{2}{v}^{2}}v\left ({\frac {\partial }{
\partial u}}y\right ){u}^{2}\!+\!\sqrt {\left (y\right )^{2}-
\left ({\frac {\partial }{\partial u}}y\right )^{2}{u}^{2}}u \left ({\frac {\partial
}{\partial v}}y\right ){v}^{2}\right ){ \frac {\partial ^{2}}{\partial u\partial
v}}y+\]\[+\sqrt {\left (y(u,v )\right )^{2}\!-\!\left ({\frac {\partial }{\partial
v}}y\right )^{2} {v}^{2}}v\left ({\frac {\partial }{\partial v}}y\right )y\!-\!
\sqrt {\left (y\right )^{2}\!-\!\left ({\frac {\partial }{\partial u} }y\right
)^{2}{u}^{2}}u\left ({\frac {\partial }{\partial u}}y\right )y=0. $$
\end{rem}

\end{document}